\RequirePackage{fix-cm}

\documentclass[twocolumn]{svjour3}          % twocolumn
\usepackage{lipsum}
\smartqed   %flush right qed marks, e.g. at end of proof
\usepackage{graphicx}
\usepackage{epstopdf}
\usepackage{amsmath}
\begin{document} \sloppy
%\lipsum[1-14]
\title{A coarse-grid incremental pressure-projection method for accelerating low Reynolds-number
incompressible flow simulations
}



\author{A. Kashefi       
}


\institute{A. Kashefi \at
              Department of Mechanical Engineering, Stanford University, Stanford, CA 94305, USA \\
  %            Tel.: +123-45-678910\\
    %          Fax: +123-45-678910\\
              \email{kashefi@stanford.edu}             \\
  %         \and
}

\date{Received: date / Accepted: date}

\maketitle

\begin{abstract}
Coarse grid projection (CGP) multigrid techniques are applicable to sets of equations that
include at least one decoupled linear elliptic equation. In CGP, the linear elliptic equation is
solved on a coarsened grid compared to the other equations, leading to savings in computations
time and complexity. One of the most important applications of CGP is when a pressure
correction scheme is used to obtain a numerical solution to the Navier-Stokes equations. In that
case there is an elliptic pressure Poisson equation. Depending on the pressure correction scheme
used, the CGP method and its performance in terms of acceleration rate and accuracy level vary.
The CGP framework has been established for non-incremental pressure projection techniques. In
this article, we apply CGP methodology for the first time to incremental pressure correction
schemes. Both standard and rotational forms of the incremental algorithms are considered. The
influence of velocity Dirichlet and natural homogenous boundary conditions in regular and
irregular domains with structured and unstructured triangular finite element meshes is
investigated.
$L^2$ norms demonstrate that the level of accuracy of the velocity and the pressure
fields is preserved for up to three levels of coarsening. For the test cases investigated, the
speedup factors range from 1.248 to 102.715.
\end{abstract}

\section{Introduction and motivation}
\label{intro}

Projection methods [1--3] are popular schemes for simulating the unsteady incompressible
Navier-Stokes equations, since the technique overcomes the saddle-point issue of the mass and
momentum conservation equations by replacing those two equations with two decoupled elliptic
ones: a nonlinear advection diffusion equation and a linear Poisson equation. Notwithstanding
this benefit, the solution of Poisson's equation is a major issue as it imposes high computational
expenses to the system [4, 5].

Since we deal with the nonlinear convection term in the momentum equation, high spatial
resolution is a key for conservation of the fidelity of the velocity field, especially for high
Reynolds numbers. On the other hand, as the Poisson equation is a linear partial differential
equation, such a refined grid resolution is not essential for its solution. Hence, an idea to
accelerate these types of simulations is to solve the nonlinear momentum equation on a fine grid
and compute the pressure Poisson equation on a corresponding coarsened grid. In 2010 Lentine
et al. [4] first proposed this multiresolution technique, called Coarse Grid Projection (CGP)
methodology, to lessen the computational cost associated with the Poisson equation for inviscid
flow simulations. In 2013 San and Staples [5] expanded CGP to the incompressible Navier-Stoke
equations (labeled ``CGPRK3''). Moreover, they applied the CGP technique to elliptic equations
of potential vorticity in quasigeostrophic ocean models [6]. In 2014 Jin and Chen [7] used CGP
for the fast fluid dynamics (FFD) models to study building airflows. In 2018 Kashefi and Staples
[8--11] presented a semi-implicit-time integration finite-element version of the CGP method (labeled
``IFE-CGP'').

In all the methods cited above, the authors [4--6, 8--11] applied CGP to the non-incremental pressure
projection scheme [1--3]. There are several limitations with this scheme which affect the
efficiency of the CGP algorithm. The performance of the CGP technique can be measured by
means of two technical parameters: speedup factor and accuracy level. For each CGP simulation,
we look for how many levels of coarsening can be performed while preserving the accuracy level
in either the velocity or the pressure field, and the associated computational speedup. The
CGPRK3 approach [5] significantly reduced the integrity of the pressure field even for one
coarsening level. In addition, a considerable reduction in the accuracy of the velocity field was
observed for two and three levels of coarsening. San and Staples [5] achieved speedup factors
ranging from roughly 2 to 42 using CGPRK3. Kashefi and Staples [9] demonstrated that IFECGP
was only able to preserve the accuracy of the pressure gradient not the pressure itself (see
e.g., Fig. 12 of Ref. [9]). Like CGPRK3, the IFE-CGP computations lost the superb fidelity of
the velocity field for more than one coarsening level. The splitting error of the non-incremental
pressure correction method [1--2] is irreducibly first-order in time with Dirichlet boundary
conditions [3]. Due to the artificial Neumann boundary conditions for the pressure, the overall
accuracy of this projection scheme is dominated by the temporal error rather than the spatial one
[3]. Hence, IFE-CGP experienced shortcomings with realistic boundary conditions. The speedup
factors of the numerical studies by IFE-CGP ranged from approximately 2 to 30.

To obviate the aforementioned problems, we implement the CGP strategy in the incremental
pressure correction schemes including the standard [3, 12] and rotational [3, 12--13] forms. Taking
this approach, the Poisson equation is solved on a coarsened grid for an intermediate variable and
not for the pressure itself. Combining incremental pressure projection methods and CGP
enhances the CGP capability in several ways. First, CGP preserves the accuracy of the velocity
and the pressure field for a high level of the Poisson equation grid coarsening and thus
remarkable speedup is reached. Second, since the incremental pressure projection scheme in
standard form has an irreducible second-order time stepping error [3], a CGP algorithm with the
standard form is improved from temporal integration point of view. Third, the incremental
pressure correction technique in rotational form overcomes the artificial layers caused by the
artificial homogenous pressure Neumann conditions [3]. Hence, a CGP method with the
rotational form inherits this feature as well. We investigate the performance of the CGP
algorithm in incremental pressure correction schemes through three standard test cases: the
Taylor-Green vortex [14] with velocity Dirichlet boundary conditions in a square, the Jobelin
vortex [12] with open boundary conditions in a square, and the Jobelin vortex [12] with velocity
Dirichlet boundary conditions in a circle.

The present work is structured as follows. The governing equations for incompressible flows and
their spatial/temporal discretizations are given in Sect. 2.1. The CGP algorithm and its
computational consideration are discussed in Sect. 2.2 and Sect. 2.3, respectively. Numerical
results are collected in Sect. 3 and conclusion is given in Sect. 4.

\section{Problem formulation}
\subsection{Governing equations}
\label{sec:2}
We consider an incompressible isothermal flow of a Newtonian fluid, which is governed by the
dimensionless form of the Navier-Stokes and continuity equations:
\begin{equation}
\bigg[\frac{\partial \textbf{\textit{u}}}{\partial t}+(\textbf{\textit{u}}\cdot \nabla)\textbf{\textit{u}}\bigg] -\frac{1}{Re} \Delta \textbf{\textit{u}} + \nabla p=\textbf{\textit{f}} \textrm{ in } V,
\end{equation}
\begin{equation}
\nabla \cdot \textbf{\textit{u}}=0 \textrm{ in } V,
\end{equation}
\begin{equation}
\textbf{\textit{u}}=\textbf{\textit{u}}_{\Gamma_D} \textrm{ on } \Gamma_D,
\end{equation}
\begin{equation}
-p\textbf{\textit{n}}+\frac{1}{Re}\nabla \textbf{\textit{u}} \cdot \textbf{\textit{n}}=\textbf{\textit{t}}_{\Gamma_N} \textrm{ on } \Gamma_N,
\end{equation}
where $\textbf{\textit{u}}$ and $p$ stand for the velocity vector and the pressure of the fluid in domain $V$, respectively. $\textbf{\textit{f}}$ represents the vector of external force and  $\textbf{\textit{t}}_{\Gamma_N}$ denotes the stress vector. $Re$ is
the Reynolds number. $\Gamma_D$ and $\Gamma_N$ respectively represent the Dirichlet and Neumann boundaries
of the domain $V$, where $\textbf{\textit{n}}$ denotes the outward unit vector normal to them. Note that there is no
overlapping between $\Gamma_D$ and $\Gamma_N$ subdomains.

Discretizing the system of equations using a second order backward differentiation formula [15]
with respect to the time variable yields to:

\begin{equation}
\begin{aligned}
\bigg[\frac{\frac{3}{2}\textbf{\textit{u}}^{n+1} - 2\textbf{\textit{u}}^n + \frac{1}{2}\textbf{\textit{u}}^{n-1}}{\delta t}+((2\textbf{\textit{u}}^n - \textbf{\textit{u}}^{n-1})\cdot \nabla)\textbf{\textit{u}}^{n+1}\bigg]\\ -\frac{1}{Re} \Delta  \textbf{\textit{u}}^{n+1} + \nabla p^{n+1}
=\textbf{\textit{f}}^{n+1} \textrm{ in } V,
\end{aligned}
\end{equation}

\begin{equation}
\nabla \cdot \textbf{\textit{u}}^{n+1}=0 \textrm{ in } V,
\end{equation}
\begin{equation}
\textbf{\textit{u}}^{n+1}=\textbf{\textit{u}}_{\Gamma_D}^{n+1} \textrm{ on } \Gamma_D,
\end{equation}
\begin{equation}
-p^{n+1}\textbf{\textit{n}}+\frac{1}{Re} \nabla \textbf{\textit{u}}^{n+1} \cdot \textbf{\textit{n}}=\textbf{\textit{t}}_{\Gamma_N}^{n+1} \textrm{ on } \Gamma_N,
\end{equation}
where $\delta t$ represents the time step. In order to obtain the numerical solution of Eqs. (5)--(8), we
utilize incremental pressure correction schemes [3]. Accordingly, at each time step $t^{n+1}$, we solve
two cascading elliptic problems: a linearized equation for the intermediate velocity field $\tilde{\textbf{\textit{u}}}^{n+1}$, and a linear Poisson's equation for an intermediate variable $\phi$. Afterwards, the end-of-step
velocity $\textbf{\textit{u}}^{n+1}$ and the pressure $p^{n+1}$ are calculated through two correction equations. The
corresponding equations are as follows:

\begin{equation}
\begin{aligned}
\bigg[\frac{\frac{3}{2}\textbf{\textit{\~u}}^{n+1} - 2\textbf{\textit{u}}^n + \frac{1}{2}\textbf{\textit{u}}^{n-1}}{\delta t}+((2\textbf{\textit{u}}^n - \textbf{\textit{u}}^{n-1})\cdot \nabla)\textbf{\textit{\~u}}^{n+1}\bigg]\\ -\frac{1}{Re} \Delta  \textbf{\textit{\~u}}^{n+1} 
=- \nabla p^{n} + \textbf{\textit{f}}^{n+1} \textrm{ in } V,
\end{aligned}
\end{equation}

\begin{equation}
\textbf{\textit{\~u}}^{n+1}=\textbf{\textit{u}}_{\Gamma_D}^{n+1} \textrm{ on } \Gamma_D,
\end{equation}

\begin{equation}
-p\textbf{\textit{n}}+\frac{1}{Re}\nabla \textbf{\textit{\~u}}^{n+1} \cdot \textbf{\textit{n}}=\textbf{\textit{t}}_{\Gamma_N}^{n+1} \textrm{ on } \Gamma_N,
\end{equation}

\begin{equation}
\Delta \phi=\frac{3}{2}\frac{1}{\delta t} \nabla \cdot \textbf{\textit{\~u}}^{n+1} \textrm{ in } V,
\end{equation}

\begin{equation}
\nabla \phi \cdot \textbf{n}=0  \textrm{ on } \Gamma_D,
\end{equation}

\begin{equation}
\phi=0  \textrm{ on } \Gamma_N,
\end{equation}

\begin{equation}
\textbf{\textit{u}}^{n+1}=\textbf{\textit{\~u}}^{n+1}-\frac{2}{3}\delta t\nabla \phi,
\end{equation}

\begin{equation}
p^{n+1}=p^{n}+\phi-\chi \frac{1}{Re} \nabla \cdot \textbf{\textit{\~u}}^{n+1},
\end{equation}
where $\chi$ is a coefficient. If $\chi=0$ the standard form of the incremental pressure correction
scheme is captured, whereas $\chi=1$ leads to the rotational form of the method.

Eqs. (9)--(16) can be spatially discretized using any desired method. Here, we use the finite element
Galerkin scheme [13] to approximate the space of velocity and pressure. Since the
projection method overcomes the saddle-point issue of Eqs. (1)--(2), satisfying the discrete
Brezzi-Babuska condition [17, 18] is not essential [13]. Hence, the piecewise linear basis function
(\textbf{P}$_1$/\textbf{P}$_1$) is implemented for the discretization of both the velocity and pressure variables. With
this in mind, the finite element form of Eqs. (9)--(16) is expressed as

\begin{equation}
\begin{aligned}
\frac{1}{\delta t}\big(\frac{3}{2}\textbf{M}_v\textrm{\~U}^{n+1}
-2\textbf{M}_v\textrm{U}^{n}+\frac{1}{2}\textbf{M}_v\textrm{U}^{n-1})+
\\
\big[\textbf{N}+\textbf{L}_v\big]\textrm{\~U}^{n+1}=-\textbf{G}\textrm{P}^{n}+\textbf{M}_v\textrm{F}^{n+1},
\end{aligned}
\end{equation}

\begin{equation}
\textbf{L}_p\Phi=\frac{3}{2}\frac{1}{\delta t}\textbf{D}\textrm{\~U}^{n+1},
\end{equation}

\begin{equation}
\textbf{M}_v\textrm{U}^{n+1}=\textbf{M}_v\textrm{\~U}^{n+1}-\frac{2}{3}\delta
t\textbf{G}\Phi,
\end{equation}

\begin{equation}
\textbf{M}_p\textrm{P}^{n+1}=\textbf{M}_p\textrm{P}^{n} + \textbf{M}_p\Phi -\chi \frac{1}{Re} \textbf{D}\textrm{\~U}^{n+1},
\end{equation}
where \textbf{M}$_v$,  \textbf{M}$_p$, \textbf{N}, \textbf{L}$_v$, \textbf{L}$_p$, \textbf{D}, and \textbf{G} indicate the matrices associated, respectively, to the velocity mass, pressure mass, nonlinear convection, velocity Laplacian, pressure Laplacian, divergence, and gradient operators. The nodal values of the intermediate variable, the intermediate velocity, the end-of-step
velocity, the forcing term, and the pressure at time $t^{n+1}$, respectively, gather in the vectors $\Phi$, {\~U}$^{n+1}$,
{U}$^{n+1}$, {F}$^{n+1}$, and {P}$^{n+1}$.

\subsection{Coarse grid projection methodology}
\label{sec:3}

The main idea of the CGP scheme is solving the Poisson equation subproblem on a coarsened
grid. Since this is the most time consuming component of the pressure-correction process, a
reduction in the degrees of freedom of the discretized Poisson equation leads to the acceleration
of these simulations. In practice, the procedure at each time step $t^{n+1}$ is as follows:
\\
(i) Obtain the intermediate velocity field data $\textrm{\~U}^{n+1}_{f}$ on a fine grid by solving the advection-diffusion equation.
\\
\\
(ii) Restrict $\textrm{\~U}^{n+1}_{f}$ to a coarsened grid to find $\textrm{\~U}^{n+1}_{c}$.
\\
\\
(iii) Solve the Poisson equation for $\Phi_c$ and set the divergence of $\textrm{\~U}^{n+1}_{c}$ as its source term.
\\
\\
(iv) Prolong the solution of the Poisson equation $\Phi_c$ to the fine grid to find $\Phi_f$.
\\
\\
(v) Correct the velocity domain on the fine grid and obtain $\textrm{U}^{n+1}_{f}$.
\\
\\
(vi) Update the pressure field on the fine grid and obtain $\textrm{P}^{n+1}_{f}$.

Geometric Multigrid (GMG) tools (see e.g., [19]) are used for the derivation of the mapping
operators. In this way, hierarchical meshes are generated by subdividing each triangular element
of a coarse grid into four triangles. Consider, for example, a coarse mesh with $N$ elements. A
fine mesh with $M$ elements is obtained by $k-$level uniform mesh refinement of the coarse grid
such that $N=4^{-k}M$. In this study, we define the restriction, $R:V_{4-l} \rightarrow V_{4}$, and prolongation, $P:V_l \rightarrow V_{l+1}$, operators for $l=$1, 2, and 3, representing mapping functions for a sequence of four nested spaces, $V_1 \subset V_2 \subset V_3 \subset V_4=V$, wherein if $V_{l+1}$ characterizes the space of a fine mesh, $V_l$ corresponds to the space of the next coarsest mesh. The principle addressed in Sect. 2.3 of Ref.[9] is followed in order to construct the matrix representation of the restriction \textbf{R}$_{\textbf{4}}^{\textbf{4-l}}$ and prolongation \textbf{P}$_{\textbf{l}}^{\textbf{l+1}}$ operators. Consider two nodes located at
$(x_f, y_f)\in V_{l+1}$ and $(x_c, y_c)\in V_l$ respectively on a fine grid and a corresponding coarsened grid. A pure injection process is used to restrict the intermediate velocity data such that $\textrm{\~U}^{n+1}_{f} (x_f, y_f)=\textrm{\~U}^{n+1}_{c} (x_c, y_c)$ if $x_f=x_c$ and $y_f=y_c$. A linear interpolation is used to prolong the intermediate pressure data such that $\Phi^{n+1}_f(x_f, y_f)= (\Phi^{n+1}_c(x_c^{'}, y_c^{'}) + \Phi^{n+1}_c(x_c^{''}, y_c^{''}))/2$ if $x_f = (x_c^{'}+x_c^{''})/2$ and $y_f = (y_c^{'}+y_c^{''})/2$. Since we utilize GMG techniques, the Laplacian ($\bar{\textbf{L}}_p$) and divergence ($\bar{\textbf{D}}$) operators of a coarsened mesh
$(V_{4-l})$ are directly derived by taking the inner products of the coarse-grid finite-element shape
functions. One may refer to Sect. 2.3 of Ref. [9] for further details.

Eqs. (21)--(26) summarize the CGP algorithm described for the incremental pressure correction
schemes. \\
1.	Calculate {\~U}$_f^{n+1}$ on $V$ by solving
\begin{equation}
\begin{aligned}
\big(\frac{3}{2}\textbf{M}_v+\delta t \textbf{N} + \delta t \textbf{L}_v\big)\textrm{\~U}_f^{n+1}=
-\delta t\textbf{G}\textrm{P}^{n}_f + \delta t\textbf{M}_v\textrm{F}^{n+1} \\ + 2\textbf{M}_v\textrm{U}_f^{n}-\frac{1}{2}\textbf{M}_v\textrm{U}_f^{n-1}.
\end{aligned}
\end{equation}
2.	Map {\~U}$_f^{n+1}$ onto $V_{4-l}$ and obtain {\~U}$_c^{n+1}$ via
\begin{equation}
\textrm{\~U}_c^{n+1}=\textbf{R}_\textbf{4}^{\textbf{4-l}}\textrm{\~U}_f^{n+1}.
\end{equation}
3.	Calculate $\Phi_c$ on $V_{4-l}$ by solving
\begin{equation}
\bar{\textbf{L}}_p\Phi_c=\frac{3}{2}\frac{1}{\delta t}\bar{\textbf{D}}\textrm{\~U}_c^{n+1}.
\end{equation}
4.	Remap $\Phi_c$ onto $V$ and obtain $\Phi_f$ via
\begin{equation}
\Phi_f=\textbf{P}_\textbf{l}^{\textbf{l+1}}\Phi_c.
\end{equation}
5.	Calculate U$_f^{n+1}$ via
\begin{equation}
\textbf{M}_v\textrm{U}_f^{n+1}=\textbf{M}_v\textrm{\~U}_f^{n+1}-\frac{2}{3}\delta
t\textbf{G}\Phi_f.
\end{equation}
6.	Calculate P$_f^{n+1}$ via
\begin{equation}
\textbf{M}_p\textrm{P}^{n+1}_f=\textbf{M}_p\textrm{P}^{n}_f+\textbf{M}_p{\Phi}_f-\chi \frac{1}{Re}\textbf{D}\textrm{\~U}^{n+1}_f.
\end{equation}
From the formulation point of view, there are two main differences between applying CGP to
non-incremental pressure correction schemes in comparison with incremental ones. First, in the
case of the non-incremental CGP process, we solve the Poisson equation for the pressure
variable $p$ on a coarsened grid, whereas in case of the incremental CGP algorithm, we solve
Poisson's equation for an intermediate variable $\phi$ on the coarsened grid. In fact, the spatial
resolution of both the velocity and pressure fields in incremental CGP simulations are kept on
the fine grid level. Second, in the incremental CGP formulation, the pressure gradient of the
previous time step \textbf{G}P$^{n+1}_f$ exists as the source term of the momentum equation (see Eq. (21)),
while the pressure does not have any contribution to the momentum equation in the nonincremental
CGP computations. We discuss the effect of these two points on the efficiency of the
CGP method in Sect. 3.

\subsection{Computational consideration}
\label{sec:4}
In the case of standard forms ($\chi=0$), one may directly solve the algebraic Eq. (16) instead of its
discretized form Eq. (20), which is computationally cheaper. We take this approach for our
numerical experiments. In the case of rotational forms ($\chi=1$), one may rewrite Eq. (20) in the
following form:
\begin{equation}
\textrm{P}^{n+1}=\textrm{P}^{n}+{\Phi}-\chi \frac{1}{Re}\textbf{M}^{-1}_p\textbf{D}\textrm{\~U}^{n+1},
\end{equation}
where \textbf{M}$^{-1}_p$ is the inverse of the lumped pressure mass matrix. Taking advantage of Eq. (27), the
necessity of inverting the consistent pressure mass matrix \textbf{M}$_p$ disappears and consequently a
more cost-effective procedure is obtained. However, our numerical results indicate more
accurate results for the pressure $p$ by solving Eq. (20). Hence, we use Eq. (20) for our
simulations.

An in-house C++ object oriented code is used. The ILU(0) preconditioned GMRES(m) algorithm
[20, 21] is employed. We use the public unstructured finite element grid generation software
Gmsh [22]. To accurately compare speedups of our simulations, we perform all calculations on a
single Intel(R) Xeon(R) processor with 2.66 GHz clock rate and 64 Gigabytes of RAM.

%\subsection{Subsection title}
%\label{sec:2}
%as required. Don't forget to give each section
%and subsection a unique label (see Sect.~\ref{sec:1}).

%Text with citations \cite{Ref3} and \cite{RefJ}.

\section{Results and discussion}
\label{sec:1}
In this section, we study three standard test cases: The Taylor-Green vortex [14] with velocity
Dirichlet boundary conditions, the Jobelin vortex with open boundary conditions (see Sect. 4.2 of
Ref. [12]), and the Jobelin vortex with Dirichlet boundary conditions (see Sect. 4.3 of Ref. [12]).
We indicate the mesh resolution of our simulations with the notation $M:N$, where $M$ denotes
the number of elements in a fine grid. If we coarsen the fine grid by $k$ levels, $N$ indicates the
number of elements of the resulting coarsened grid.

To save space, we mark the implimentation of CGP with the non-incremental pressure correction
scheme by ``NCGP,'' CGP with the standard incremental pressure correction technique by
``SCGP,'' and CGP with the rotational pressure correction method by ``RCGP.''

\subsection{Taylor-Green vortex with velocity Dirichlet boundary conditions}
\label{sec:2}
The concern of this section is to investigate the effects of velocity Dirichlet boundary conditions
on the performance of the SCGP and RCGP implementations of the method.

The velocity field of the two dimensional Taylor-Green vortex [14] is given by:
\begin{equation}
u(x,y,t)=-\cos(2\pi x)\sin(2\pi y)\exp(-8\pi^2 t/Re),
\end{equation}
\begin{equation}
v(x,y,t)=\sin(2\pi x)\cos(2\pi y)\exp(-8\pi^2 t/Re).
\end{equation}
And the pressure field is given by:
\begin{equation}
p(x,y,t)=-\frac{\cos(4\pi x)+\cos(4\pi y)}{4}\exp(-16\pi^2 t/Re).
\end{equation}

We impose the exact solution of Eqs. (28)--(29) on the velocity domain boundaries while we solve the
Poisson equation with homogenous artificial Neumann boundary conditions (see Eq. (13)). The
numerical studies are executed until time $t=1$.

We simulate the Taylor-Green vortex [14] for a Reynolds number of $Re=10$ in the
computational domain $V:=[0, 1]\times[0, 1]$ with different grid resolutions. The simulations are run
with a constant time step $\delta t=0.00125$.

The discrete norms of the velocity, the pressure, and the pressure gradient fields are tabulated
respectively in Tables 1--3 for different mesh resolutions for both the standard and the rotational
forms at time $t=1$.

\begin{table*}
\centering
\caption{Velocity error norms for different grid resolutions of the Taylor-Green vortex simulation
at $t=1$. $M:N$ represents the grid resolution of the advection-diffusion solver ($M$ elements)
and Poisson's equation ($N$ elements).}
\label{tab:1}   
\begin{tabular}{llllll}
\hline\noalign{\smallskip}
 &  & Standard Form & &  Rotational Form & \\
\hline\noalign{\smallskip}
$k$ & Resolution & $\| \textbf{\textit{u}}\|_{L^\infty(V)}$ & $\| \textbf{\textit{u}}\|_{L^2(V)}$ &  $\| \textbf{\textit{u}}\|_{L^\infty(V)}$ & $\| \textbf{\textit{u}}\|_{L^2(V)}$\\
\noalign{\smallskip}\hline\noalign{\smallskip}
0 & 65536:65536 & 1.90075E$-$6 & 1.55734E$-$6 & 1.90075E$-$6 & 1.55734E$-$6 \\
1 & 65536:16384 & 1.90075E$-$6 & 1.55734E$-$6 & 1.90075E$-$6& 1.55735E$-$6 \\
2 & 65536:4096 & 1.90075E$-$6 & 1.55735E$-$6 & 1.90075E$-$6 & 1.55735E$-$6 \\
3 & 65536:1024 & 1.90075E$-$6 & 1.55736E$-$6 & 1.90075E$-$6 & 1.55737E$-$6\\
\noalign{\smallskip}\hline\noalign{\smallskip}
0 & 16384:16384 & 7.60304E$-$6 & 6.22685E$-$6 & 7.60304E$-$6 & 6.22685E$-$6 \\
1 & 16384:4096 & 7.60304E$-$6 & 6.22686E$-$6 & 7.60304E$-$6 & 6.22686E$-$6 \\
2 & 16384:1024 & 7.60304E$-$6 &6.22688E$-$6 & 7.60304E$-$6 & 6.22687E$-$6 \\
\noalign{\smallskip}\hline\noalign{\smallskip}
0 & 4096:4096 & 3.04127E$-$5 & 2.48677E$-$5 & 3.04127E$-$5 & 2.48677E$-$5 \\
1 & 4096:1024 & 3.04127E$-$5  & 2.48678E$-$5 & 3.04127E$-$5 & 2.48677E$-$5 \\
\noalign{\smallskip}\hline\noalign{\smallskip}
0 & 1024:1024 & 0.00012166 & 9.88459E$-$7 & 0.00012166  & 9.88459E$-$7 \\
\noalign{\smallskip}\hline
\end{tabular}
\end{table*}

\begin{table*}
\centering
\caption{Pressure error norms for different grid resolutions of the Taylor-Green vortex simulation
at $t=1$. $M:N$ represents the grid resolution of the advection-diffusion solver ($M$ elements)
and Poisson's equation ($N$ elements).}
\label{tab:2}   
\begin{tabular}{llllll}
\hline\noalign{\smallskip}
 &  & Standard Form & &  Rotational Form & \\
\hline\noalign{\smallskip}
$k$ & Resolution & $\| {\textit{p}}\|_{L^\infty(V)}$ & $\| {\textit{p}}\|_{L^2(V)}$ &  $\| {\textit{p}}\|_{L^\infty(V)}$ & $\| {\textit{p}}\|_{L^2(V)}$\\
\noalign{\smallskip}\hline\noalign{\smallskip}
0 & 65536:65536 & 3.63539E$-$06 & 2.06204E$-$06 & 3.63539E$-$06 & 2.06172E$-$06\\
1 & 65536:16384 & 3.63539E$-$06 & 2.06205E$-$06 & 3.63539E$-$06 & 2.06187E$-$06 \\
2 & 65536:4096 & 3.63539E$-$06 & 2.06207E$-$06 & 3.63539E$-$06 & 2.06194E$-$06\\
3 & 65536:1024 & 0.000176743 & 0.000136857 & 3.63539E$-$06 & 2.06199E$-$06 \\
\noalign{\smallskip}\hline\noalign{\smallskip}
0 & 16384:16384 & 1.43317E$-$05 & 8.17426E$-$06 &  1.43317E$-$05 & 8.17290E$-$06 \\
1 & 16384:4096 & 1.43317E$-$05 & 8.17428E$-$06 &  1.43317E$-$05 & 8.17352E$-$06 \\
2 & 16384:1024 & 0.00199501 & 0.00152358 &  1.43317E$-$05 & 8.17385E$-$06 \\
\noalign{\smallskip}\hline\noalign{\smallskip}
0 & 4096:4096 & 5.14363E$-$05 & 3.25573E$-$05 & 5.14363E-05 & 3.25546E$-$5 \\
1 & 4096:1024 & 5.14363E$-$05  & 3.25574E$-$05 & 5.14366E-05 & 3.25523E$-$5 \\
\noalign{\smallskip}\hline\noalign{\smallskip}
0 & 1024:1024 & 0.000216977 & 0.000129742 & 0.000216977  & 0.000129722 \\
\noalign{\smallskip}\hline
\end{tabular}
\end{table*}

\begin{table*}
\centering
\caption{Pressure gradient error norms for different grid resolutions of the Taylor-Green vortex
simulation at $t=1$. $M:N$ represents the grid resolution of the advection-diffusion solver ($M$
elements) and Poisson's equation ($N$ elements).}
\label{tab:3}   
\begin{tabular}{llllll}
\hline\noalign{\smallskip}
 &  & Standard Form & &  Rotational Form & \\
\hline\noalign{\smallskip}
$k$ & Resolution & $\| \textbf{G}$P$\|_{L^\infty(V)}$ & $\| \textbf{G}$P$\|_{L^2(V)}$ &  $\| \textbf{G}$P$\|_{L^\infty(V)}$ & $\| \textbf{G}$P$\|_{L^2(V)}$\\
\noalign{\smallskip}\hline\noalign{\smallskip}
0 & 65536:65536 & 9.58885E$-$14 & 1.02348E$-$14 & 3.77312E$-$13 & 3.55036E$-$14 \\
1 & 65536:16384 & 3.44286E$-$13 & 8.93675E$-$14 & 4.48572E$-$13 & 3.86897E$-$14 \\
2 & 65536:4096 & 4.26302E$-$13 & 1.99537E$-$13 & 4.49496E$-$13 & 3.87436E$-$14 \\
3 & 65536:1024 & 3.53933E$-$12 & 1.72439E$-$12 & 4.53791E$-$13 & 3.90375E$-$14\\
\noalign{\smallskip}\hline\noalign{\smallskip}
0 & 16384:16384 & 2.11731E$-$12 & 3.21126E$-$13 & 5.74009E$-$12 & 7.38968E$-$13 \\
1 & 16384:4096 & 8.57518E$-$12 & 3.56109E$-$12 & 7.08534E$-$12 & 9.19657E$-$13 \\
2 & 16384:1024 & 3.05404E$-$11 & 1.40611E$-$11 & 7.13954E$-$12 & 9.22436E$-$13 \\
\noalign{\smallskip}\hline\noalign{\smallskip}
0 & 4096:4096 & 8.99248E$-$11 & 1.56495E$-$11 & 3.21784E$-$11 & 5.12380E$-$12 \\
1 & 4096:1024 & 9.00494E$-$11  & 5.90353E$-$11 & 6.33028E$-$11 & 1.13387E$-$11 \\
\noalign{\smallskip}\hline\noalign{\smallskip}
0 & 1024:1024 & 3.80987E$-$10 & 1.15607E$-$10 & 3.31517E$-$10 & 9.91869E$-$11 \\
\noalign{\smallskip}\hline
\end{tabular}
\end{table*}

\begin{table*}
\centering
\caption{CPU times and relative speedups for different grid resolutions of the Taylor-Green
vortex simulation at $t=1$. $M:N$ represents the grid resolution of the advection-diffusion solver
($M$ elements) and Poisson's equation ($N$ elements).}
\label{tab:4}   
\begin{tabular}{llllll}
\hline\noalign{\smallskip}
 &  & Standard Form & &  Rotational Form & \\
\hline\noalign{\smallskip}
$k$ & Resolution & CPU time (s) & Speedup & CPU time (s) & Speedup\\
\noalign{\smallskip}\hline\noalign{\smallskip}
0 & 65536:65536 & 10372.90 & 1.000 & 10446.00 & 1.000\\
1 & 65536:16384 & 8305.65 & 1.248 & 8263.63 & 1.264 \\
2 & 65536:4096 & 7312.59 & 1.418 & 7256.50 & 1.439 \\
3 & 65536:1024 & 7234.28 & 1.433 & 7187.89 & 1.453 \\
\noalign{\smallskip}\hline\noalign{\smallskip}
0 & 16384:16384 & 724.61 & 1.000 &  736.07 & 1.000 \\
1 & 16384:4096 & 548.19 & 1.321 & 548.02 & 1.343 \\
2 & 16384:1024 & 478.97 & 1.512 &  470.29 & 1.565 \\
\noalign{\smallskip}\hline\noalign{\smallskip}
0 & 4096:4096 & 61.31 & 1.000 & 63.19 & 1.000 \\
1 & 4096:1024 & 38.58  & 1.589 & 37.67 & 1.677 \\
\noalign{\smallskip}\hline\noalign{\smallskip}
0 & 1024:1024 & 6.05 & 1.000 & 1.050  & 1.000 \\
\noalign{\smallskip}\hline
\end{tabular}
\end{table*}

As far as the velocity error norms are concerned, both the SCGP and RCGP approaches preserve
the accuracy level of the field for all mesh coarsening levels that we consider. For instance, the
infinity and $L^2$ norms calculated for full fine (65536:65536), $k=1$ (65536:16384), $k=2$
(65536:4096), and $k=3$ (65536:1024) computations are approximately identical.

For the pressure, RCGP is more successful than SCGP in maintaining the pressure field accuracy
for two and three coarsening levels. For example, consider the standard fine scale 65536:65536
gird resolution $(k=0)$. The associated $L^2$ norms are equal to 2.06204E-06 and 2.06172E-06,
respectively, using the standard and rotational incremental pressure projection schemes. By
choosing $k=3$ (65536:1024), the $L^2$ norms change to 0.000136857 and 2.06199E-06,
respectively, for SCGP and RCGP, indicating 6536.971\% and 0.013\% error increase with
reference to the regular fine scale $(k=0)$ computations. This trend also occurs when we
compare the resulting data of the pure fine 16384:16384 spatial resolution $(k=0)$ with the CGP
16384:1024 grid resolution $(k=2)$. Here we illustrate the cause. Looking at Eq. (26), the end-of-step pressure P$^{n+1}_f$
is corrected by divergence of the intermediate velocity field $\frac{1}{Re}$\textbf{D}\~{U}$_f^{n+1}$ in
the rotational form formulation, while this term is neglected in standard form computations. The
intermediate velocity field \~{U}$_f^{n+1}$ is calculated on a fine grid, in contrast with the intermediate
pressure variable ${\Phi}_f$, which is prolonged from the corresponding coarsened grid data ${\Phi}_c$. Thus,
for high Poisson grid coarsening levels, when ${\Phi}_c$, and consequently ${\Phi}_f$, includes relatively
large errors, the additional divergence of the intermediate velocity field term can mitigate these
errors in the pressure field.

Concerning the pressure gradient, we observe similar trends between the pressure and the
pressure gradient $L^2$ norms for SCGP and RCGP. For example, the pressure gradient $L^2$ norms
for SCGP for $k=1$ (65536:16384), $k=2$ (65536:4096), and $k=3$ (65536:1024), respectively,
imply 773.173\%, 1849.593\%, and 16748.301\% error increases, whereas for RCGP they imply
8.974\%, 9.125\%, and 9.953\% error increases, all with reference to $k=0$ (65536:65536). The
data indicate the higher capacity of RCGP for preserving the accuracy of the pressure gradient
field.

Note that San and Staples [5] have also studied this problem at Reynolds number of $Re=10.0$ using
NCGP. However, their method totally lost the accuracy of the pressure field even after one level
coarsening. According to Table 3 of Ref. [5], the velocity $L^2$ norms for $k=1$, $k=2$, and $k=3$,
respectively, implied 1.141\%, 218.483\%, and 2465.824\% error increases, with reference to
$k=0$.

The corresponding CPU times and acceleration rates are tabulated in Table 4. The speedup
factors achieved range from 1.248 to 1.677. For each spatial resolution, the rotational form
demonstrates slightly higher speedup factors in comparison with the standard forms.

\subsection{Jobelin vortex with open boundary conditions}
\label{sec:3}
To study the capability of the proposed CGP framework in the presence of open boundary
conditions, we analyze the vortex introduced by Jobelin et al. [12]. Based on it, the forcing term of
the Navier-Stokes equation is adjusted for the divergence free velocity field
\begin{equation}
u(x, y, t)=\sin(x)\sin(y+t),
\end{equation}
\begin{equation}
v(x, y, t)=\cos(x)\cos(y+t),
\end{equation}
and an arbitrary pressure field
\begin{equation}
p(x, y, t)=\cos(x)\sin(y+t).
\end{equation}

Jobelin et al. [12] considered this vortex for a Stokes flow simulation, while we consider the
nonlinear convection term of the Navier-Stokes equation in the present work. A Reynolds
number of $Re=10$ is used. The computational domain is set to $V:=[0, 1]\times[0, 1]$. Homogenous
natural Neumann conditions
\begin{equation}
-p\textbf{\textit{n}}+\frac{1}{Re} \nabla \textbf{\textit{u}} \cdot \textbf{\textit{n}}=0,
\end{equation}
are enforced at the y-axis, while velocity Dirichlet boundary conditions are imposed at the
remaining boundaries. The time step is chosen to be $\delta t=0.01$.

Velocity, pressure, and pressure gradient error norms are tabulated in Table 5 and Table 6,
respectively, for the SCGP and NCGP computations for several spatial resolutions at real time
$t=1$. For all levels of coarsening, SCGP keeps the level of accuracy of velocity and pressure
fields the same as the output data with regular simulations ($k=0$). For instance, the $L^2$ norms
computed on the 16384:1024 spatial resolution indicate only a 0.163\% and 0.019\% reduction,
respectively, in the accuracy level for the velocity and pressure fields with reference to the full
fine scale simulations. And, more importantly, they are two and one orders of magnitude more
accurate, respectively, in comparison with the velocity and pressure fields obtained from the full
coarse scale simulation performed with 1024: 1024 spatial resolution.

\begin{table*}
\centering
\caption{Error norms and relative speedups for different grid resolutions of the Jobelin vortex
problem with open boundary conditions using the incremental projection method (standard form)
at $t=1$. $M:N$ represents the grid resolution of the advection-diffusion solver ($M$ elements)
and Poisson's equation ($N$ elements).}
\label{tab:5}   
\begin{tabular}{llllll}
\hline\noalign{\smallskip}
$k$ & Resolution & $\| \textbf{\textit{u}}\|_{L^2(V)}$ & $\| p\|_{L^2(V)}$ & $\| \textbf{G}$P$\|_{L^2(V)}$ & Speedup\\
\noalign{\smallskip}\hline\noalign{\smallskip}
0 & 16384:16384 & 6.44488E$-$7 & 1.51654E$-$5 & 2.20803E$-$9 & 1.000 \\
1 & 16384:4096 & 6.44964E$-$7 & 1.51663E$-$5 & 2.21058E$-$9 & 3.179 \\
2 & 16384:1024 & 6.45539E$-$7 & 1.51684E$-$5 & 2.21401E$-$9 & 3.943 \\
\noalign{\smallskip}\hline\noalign{\smallskip}
0 & 4096:4096 & 4.43303E$-$6 & 5.68172E$-$5 & 3.15348E$-$8 & 1.000 \\
1 & 4096:1024 & 4.51250E$-$6  & 5.68981E$-$5 & 3.84971E$-$8 & 3.686 \\
\noalign{\smallskip}\hline\noalign{\smallskip}
0 & 1024:1024 & 2.62312E$-$5 & 0.000280331 & 4.50577E$-$7 & 1.000 \\
\noalign{\smallskip}\hline
\end{tabular}
\end{table*}

\begin{table*}
\centering
\caption{Error norms and relative speedups for different grid resolutions of the Jobelin vortex
problem with open boundary conditions using the non-incremental projection method at $t=1$.
$M:N$ represents the grid resolution of the advection-diffusion solver ($M$ elements) and
Poisson's equation ($N$ elements).}
\label{tab:6}   
\begin{tabular}{llllll}
\hline\noalign{\smallskip}
$k$ & Resolution & $\| \textbf{\textit{u}}\|_{L^2(V)}$ & $\| p\|_{L^2(V)}$ & $\| \textbf{G}$P$\|_{L^2(V)}$ & Speedup\\
\noalign{\smallskip}\hline\noalign{\smallskip}
0 & 16384:16384 & 1.39996E$-$6 & 1.53335E$-$5 & 2.71414E$-$9 & 1.000 \\
1 & 16384:4096 & 1.40080E$-$6 & 1.53402E$-$5 & 3.19550E$-$9 & 2.986 \\
2 & 16384:1024 & 1.40182E$-$6 & 1.53457E$-$5 & 3.44549E$-$9 & 3.693 \\
\noalign{\smallskip}\hline\noalign{\smallskip}
0 & 4096:4096 &4.68969E$-$6 & 6.16750E$-$5 & 3.84971E$-$8 & 1.000 \\
1 & 4096:1024 & 4.70228E$-$6  & 6.17718E$-$5 & 6.29129E$-$8 & 3.798 \\
\noalign{\smallskip}\hline\noalign{\smallskip}
0 & 1024:1024 & 1.88151E$-$5 & 0.000247378 & 1.27147E$-$6 & 1.000 \\
\noalign{\smallskip}\hline
\end{tabular}
\end{table*}

Compared to NCGP, SCGP performs noticeably more robustly in order to preserve the pressure
gradient accuracy. According to the data presented in Table 5, the pressure gradient $L^2$ norm
$||\textbf{G}$P$||_{L^2(V)}$ for $k=1$ (4096:1024) shows a 22.078\% error in comparison with $k=0$ (4096:4096);
however, this measurement is equal to 63.422\% for the NCGP computations. Based upon
Kashefi and Staples [9], the CGP methodology achieves higher speedup factors in the presence
of stress-free conditions compared to velocity Dirichlet boundary conditions. Here, our
numerical experiments illustrate similar behaviors. While the maximum speedup factor found for
two levels of coarsening $k=2$ in Sect. 3.1 is 1.565, this quantity is 3.943 in the current
section.

Similar to the Taylor-Green vortex problem, we do not observe a significant difference between
the SCGP and RCGP outputs. Thus in order to save space, we do not present the results of the
RCGP simulations.

\subsection{Jobelin vortex with Dirichlet boundary conditions}
\label{sec:4}

So far we have investigated the CGP scheme in simple square domains with structured grids. The
main goal of this section is an examination of the CGP framework in a more challenging
geometry with unstructured triangular meshes. To this purpose, we consider another vortex used
by Jobelin et al. [12] such that the velocity and pressure fields for an incompressible flow read:
\begin{equation}
u(x, y, t)=\sin(x+t)\sin(y+t),
\end{equation}
\begin{equation}
v(x, y, t)=\cos(x+t)\cos(y+t),
\end{equation}
\begin{equation}
p(x, y, t)=\cos(x-y+t),
\end{equation}
with a forcing term to balance the Navier-Stokes equations. Note that Jobelin et al. [12] performed
this simulation for Stokes flow, whereas we consider the nonlinear convection term as well. The
computational domain is a circle $V:=\{(x, y)|x^2+y^2<0.25\}$. The problem geometry is
exhibited in Fig. 1 and details of the mesh are described. The computational domain uses
velocity Dirichlet boundary conditions and consequently artificial pressure homogenous
Neumann boundary conditions. A Reynolds number of $Re=10$ is utilized. The time step chosen
for these simulations is $\delta t=0.01$.

\begin{figure*}
\centering
\includegraphics[width=185 mm]{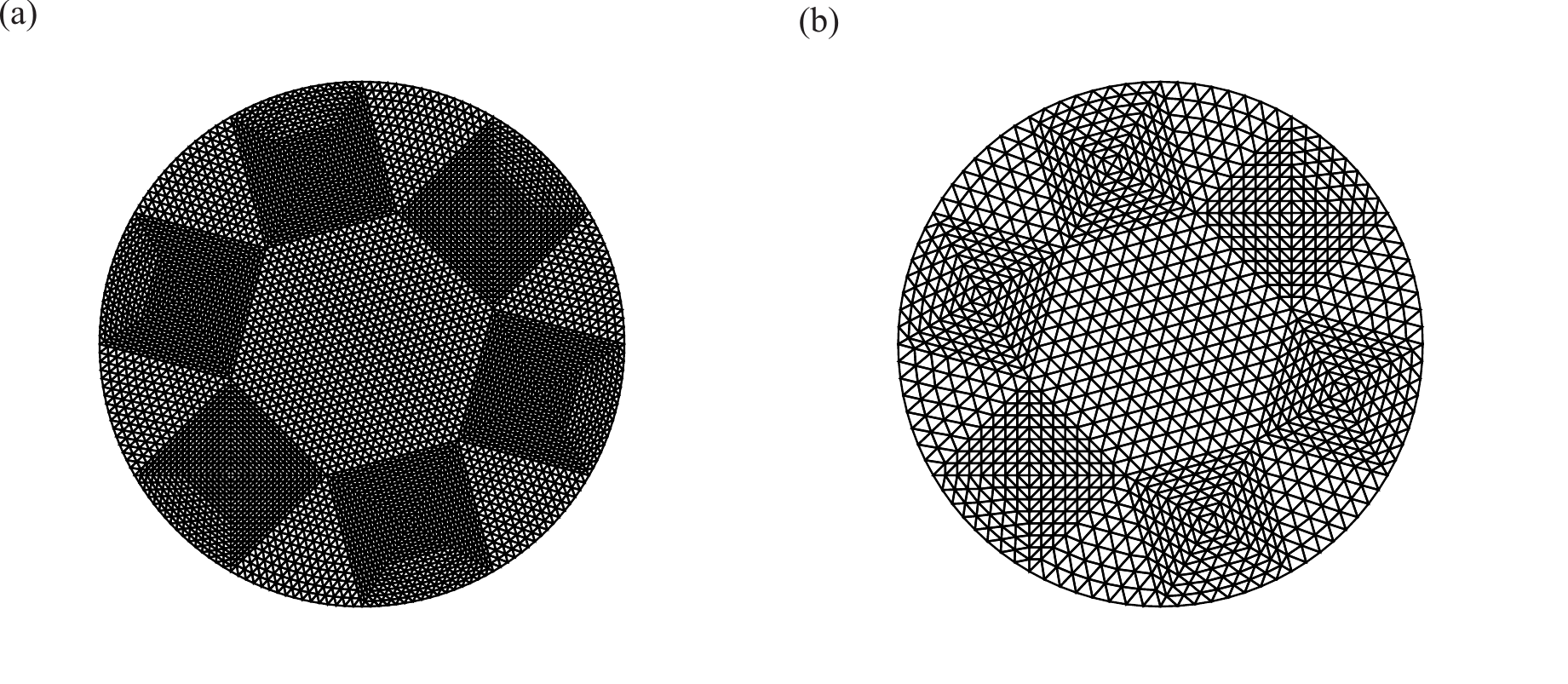}
\caption{The triangular finite element meshes used for solving Poisson's equation in the simulation
of Jobelin vortex with a Dirichlet boundary condition. \textbf{a} After one level coarsening $(k=1)$, 4705 nodes and 9216 elements; \textbf{b} After two levels coarsening $(k=2)$, 1201 nodes and 2304 elements.}
\label{fig:1} 
\end{figure*}

Tables 7--8 list the discrete error norms for the velocity, pressure, and pressure gradient fields as
well as speedup factors, respectively, for RCGP and NCGP at time $t=­1$. Considering the
36864:36864 grid resolution, after two levels $(k=2)$ of the Poisson grid coarsening, the
minimum speedup gained is equal to 3.943 and belongs to NCGP, whereas the maximum
speedup achieved is equal to 102.715 and occurs in RCGP. To more precisely discuss the
speedup factors, the relevant quantities are reported in detail. The computational times for the
performed simulations using RCGP are: 61044.8, 13281.0, and 594.31, respectively for $k=0$
(36864:36864), $k=1$ (36864:9216), and $k=2$ (36864:2304), while the same simulation using
NCGP takes: 60831.7, 27414.0, and 15427.7, respectively for $k=0$, $k=1$, and $k=2$. On the
other hand, the computational cost devoted to the Poisson equation solver in the RCGP scheme
are: 60839.9, 13076.2, and 389.28, respectively for $k=0$, $k=1$, and $k=2$, while obtaining the
solution of Poisson's equation performed by the NCGP method takes: 60626.4, 26991.2, and
14105.3, respectively, for $k=0$, $k=1$, and $k=2$. Even in unstructured grids, the RCGP system
keeps the accuracy of the pressure field to an excellent degree, as can be seen from Table 7.
Interestingly, the computational cost paid to this goal becomes inexpensive and high saving in
CPU time is gained. The NCGP tool, in contrast, preserves the accuracy of the pressure and
velocity fields in a lower order and with lower speedups. A visual demonstration of this
interpretation is displayed in Figs. 2--3.

\begin{figure*}
\centering
\includegraphics[width=185 mm]{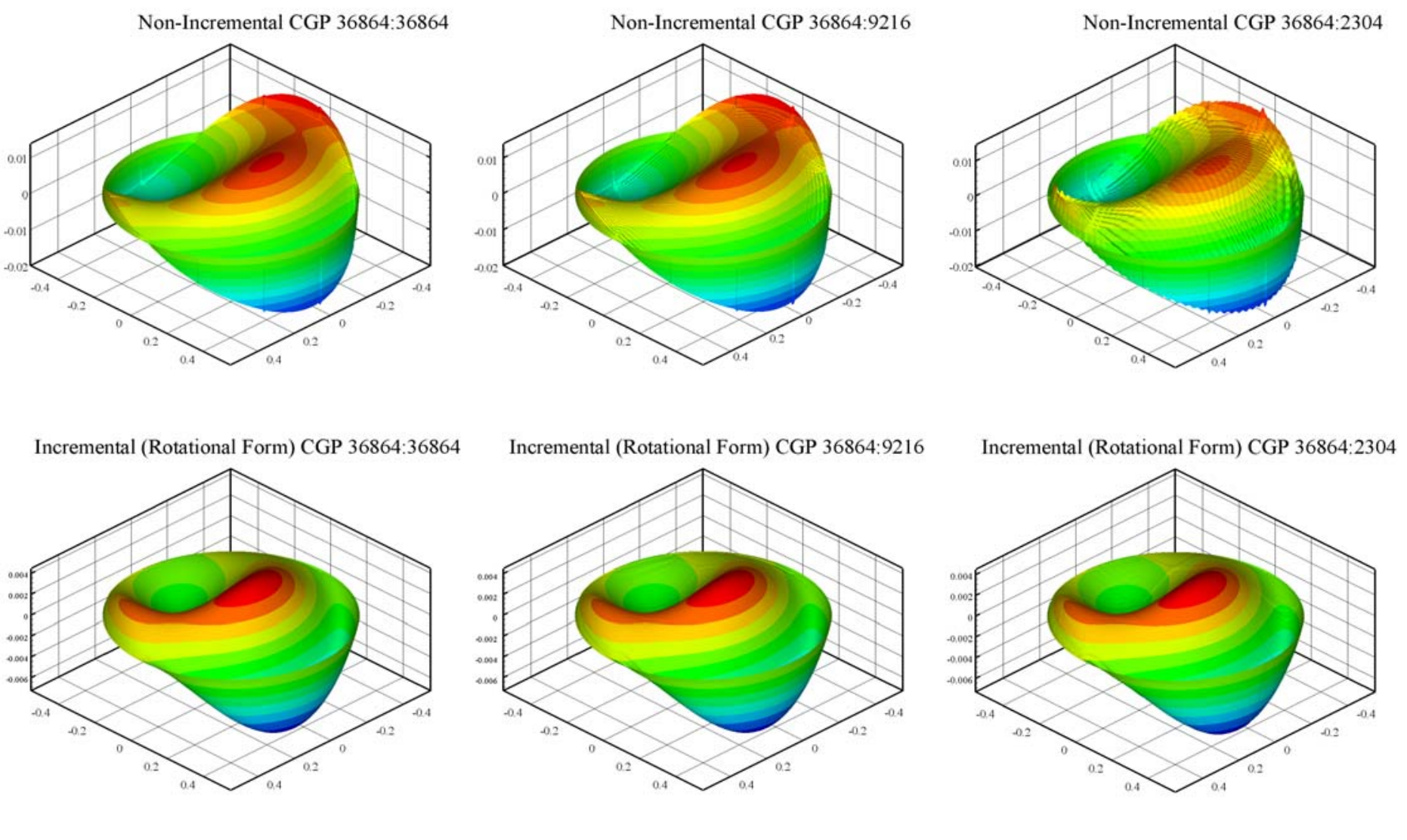}
\caption{Distribution of point-wise velocity error for Jobelin vortex problem with Dirichlet
boundary conditions at $t=1$. Labels in the form of $M:N$ indicate the grid resolution of the
advection-diffusion solver, $M$ elements, and the Poisson equation, $N$ elements.}
\label{fig:2} 
\end{figure*}

\begin{figure*}
\centering
\includegraphics[width=185 mm]{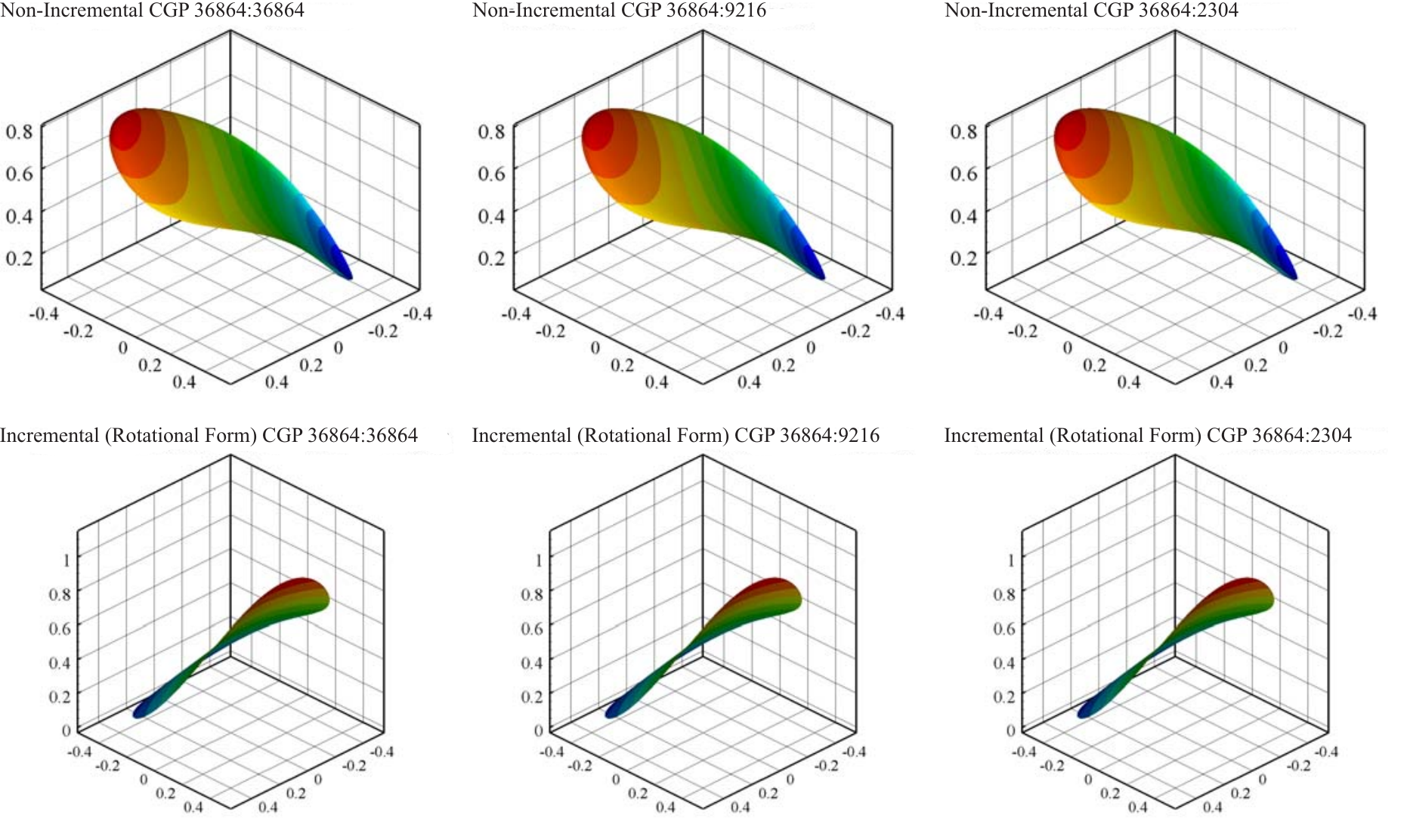}
\caption{Distribution of point-wise pressure error for Jobelin vortex problem with Dirichlet
boundary conditions at $t=1$. Labels in the form of $M:N$ indicate the grid resolution of the
advection-diffusion solver, $M$ elements, and the Poisson equation, $N$ elements.}
\label{fig:3} 
\end{figure*}

\begin{table*}
\centering
\caption{Error norms and relative speedups for different grid resolutions of the Jobelin vortex
problem with Dirichlet boundary conditions using the incremental projection method (rotational
form) at $t=1$. $M:N$ represents the grid resolution of the advection-diffusion solver ($M$ elements) and Poisson's equation ($N$ elements).}
\label{tab:7}   
\begin{tabular}{llllll}
\hline\noalign{\smallskip}
$k$ & Resolution & $\| \textbf{\textit{u}}\|_{L^2(V)}$ & $\| p\|_{L^2(V)}$ & $\| \textbf{G}$P$\|_{L^2(V)}$ & Speedup\\
\noalign{\smallskip}\hline\noalign{\smallskip}
0 & 36864:36864 & 1.97531E$-$8 & 5.19843E$-$6 & 2.37525E$-$10 & 1.000 \\
1 & 36864:9216 & 1.97531E$-$8 & 5.19843E$-$6 & 2.37525E$-$10 & 4.596 \\
2 & 36864:2304 & 1.97531E$-$8 & 5.19843E$-$6 & 2.37525E$-$10 & 102.715 \\
\noalign{\smallskip}\hline\noalign{\smallskip}
0 & 9216:9216 & 8.07724E$-$8 & 2.18577E$-$5 & 4.83658E$-$9 & 1.000 \\
1 & 9216:2304 & 8.08305E$-$8  & 2.18579E$-$5 & 4.83861E$-$9 & 2.155 \\
\noalign{\smallskip}\hline\noalign{\smallskip}
0 & 2304: 2304 & 3.16102E$-$7 & 8.62595E$-$5 & 7.51999E$-$8 & 1.000 \\
\noalign{\smallskip}\hline
\end{tabular}
\end{table*}

\begin{table*}
\centering
\caption{Error norms and relative speedups for different grid resolutions of the Jobelin vortex
problem with Dirichlet boundary conditions using the non-incremental projection method at
$t=1$. $M:N$ represents the grid resolution of the advection-diffusion solver ($M$ elements) and
Poisson's equation ($N$ elements).}
\label{tab:8}   
\begin{tabular}{llllll}
\hline\noalign{\smallskip}
$k$ & Resolution & $\| \textbf{\textit{u}}\|_{L^2(V)}$ & $\| p\|_{L^2(V)}$ & $\| \textbf{G}$P$\|_{L^2(V)}$ & Speedup\\
\noalign{\smallskip}\hline\noalign{\smallskip}
0 & 36864:36864 & 5.52553E$-$8 & 5.42703E$-$6 & 3.06539E$-$10 & 1.000 \\
1 & 36864:9216 & 5.53430E$-$8 & 5.42704E$-$6 & 3.06676E$-$10 & 2.219 \\
2 & 36864:2304 & 5.56868E$-$8 & 5.42706E$-$6 & 3.06683E$-$10 & 3.943 \\
\noalign{\smallskip}\hline\noalign{\smallskip}
0 & 9216:9216 & 2.19621E$-$7 & 2.34661E$-$5 & 1.28570E$-$8 & 1.000 \\
1 & 9216:2304 & 2.20984E$-$7 & 2.34663E$-$5 & 1.28648E$-$8 & 2.056 \\
\noalign{\smallskip}\hline\noalign{\smallskip}
0 & 2304: 2304 & 8.68060E$-$7 & 0.000110098 & 4.05365E$-$7 & 1.000 \\
\noalign{\smallskip}\hline
\end{tabular}
\end{table*}

Figure 2 and Figure 3 show the associated point-wise error distributions, respectively, for the
velocity and pressure variables using the NCGP and RCGP simulations. The general resultant
patterns of point-wise error distribution of NCGP and RCGP over the velocity domains are
identical. However, NCGP calculations lead to higher infinity norms in comparison with RCGP.
Moreover, the RCGP procedure produces identical velocity noise patterns for $k=0$, $k=1$, and
$k=2$. As shown in Fig. 3, the point-wise error distribution pattern of NCGP over the pressure
domain is completely different in comparison with those executed by RCGP. As depicted in Fig.2, because the NCGP module is disable to remove resulting artificial layers from the artificial
Neumann pressure boundary conditions, the maximum velocity noise is observed on its circular
domain boundaries, while these layers disappear in velocity domains simulated by RCGP for all
the presented resolutions.

It is worthwhile to note that SCGP is also successful in terms of accuracy and speedup levels.
However, its performance is similar to RCGP from the both aspects and that is why we only
presented the results computed by RCGP in this section.

\section{Conclusions and future directions}
\label{sec:1}

The contribution of the CGP methodology to pressure correction schemes is to accelerate the computations while preserving the accuracy of the pressure and velocity fields by evolving the nonlinear advection-diffusion equation on a fine grid and solving the linear Poisson equation on a corresponding coarsened grid. For the first time in this article, a CGP mechanism is
implemented in standard/rotational incremental pressure correctio methods. Here, Poisson's equation is solved on a coarsened mesh for an intermediate variable related to the pressure field. Hence, in contrast with the non incremental procedure, the resolution of the pressure field remains unchanged.

Three different standard test cases were solved in order to examine the performance of the
proposed CGP technique: The Taylor-Green vortex with velocity Dirichlet boundary conditions
[14], the Jobelin vortex with open boundary conditions [12], and the Jobelin vortex with Dirichlet
boundary conditions [12]. The speedup factors ranged from 1.248 to 102.715. We observed the
minimum speedup in the Taylor-Green vortex with Dirichlet boundary conditions [14] with the
standard form of the incremental pressure correction scheme, while the maximum speedup
belonged to the Jobelin vortex with Dirichlet boundary conditions [12] with the rotational form.
 
In terms of the accuracy level, generally the velocity, pressure, and pressure gradient fields, for
one, two, and three Poisson grid coarsening levels maintained excellent agreement with those
performed on full fine scale grid resolutions. In the presence of open boundary conditions, the
coarse-grid-projection incremental form of pressure correction schemes obtained velocity and
pressure norms approximately identical to those computed using full fine scale simulations. For
velocity Dirichlet boundary conditions as well as irregular physical domains, the coarse-grid projection
incremental form of pressure correction methods achieved significant speedup factors
while preserving the accuracy of both the velocity and pressure fields.

In this article, we used the method of manufactured solutions. We considered a low Reynolds
number of $Re=10$. Generally, the SCGP and RCGP techniques were prosperous. Note that
there is an important difference between the non-incremental and incremental correction
formulations. In the incremental techniques, the Poisson equation is solved for an intermediate
variable relevant to the pressure. This feature enables the CGP method to preserve the pressure
accuracy level high even for three Poisson grid coarsening levels, whereas this precision is not
obtained in the incorporation of the CGP mechanism and the non-incremental pressure projection
method. On the other hand, incremental forms force the pressure gradient term to the momentum
balance. The pressure gradient coming from CGP still experiences artificial fluctuations.

Although the magnitude of these fluctuations is low, we need to filter them at high-Reynolds number
implementations of RCGP and SCGP methodologies. Thus, designing efficient filters to
reach this goal is the topic of our future research.

Another objective of a future study is to apply a CGP method to the Nodal Discontinuous
Galerkin (NDG) [23] spatial discretization scheme. From a grid resolution point of view, in the
NDG approach the polynomial order of an element demonstrates its grid resolution. Hence,
coarsening a mesh can take a new shape. Instead of decreasing the number of elements, one may
decrease the polynomial order of the discretized space. In this way, incorporation of the CGP
methodology and the NDG scheme for incompressible flow simulations means solving the
momentum balance and the Poisson equations on grids with the same number of elements but
with different polynomial orders for each mesh. Accordingly, the advection-diffusion grid takes
higher order polynomials in comparison with the Poisson equation one. In this case, defining
novel restriction, prolongation, divergence and Laplacian operators as well as designing efficient
data structures for nodal connectivity between the advection-diffusion and the pressure grids
should be investigated.

%and subsection a unique label (see Sect.~\bibitem{Ref31}).
%\paragraph{Paragraph headings} Use paragraph headings as needed.

\begin{acknowledgements}
AK wishes to thank Dr. Peter Minev for helpful discussions.
\end{acknowledgements}



\end{document}